# Boosting the Self-driven Properties of 2D Photodetectors through Synergistic Asymmetrical Effects

*Yihong Sun, Jiefei Zhu, Yingjie Luo, Jiwei Chen, Yueyi Sun, Min Zhang, Cary Y. Yang, and Changjian Zhou\**

Y. Sun, Y. Luo, J. Chen, Y. Sun, C. Zhou
School of Microelectronics
South China University of Technology
Guangzhou 511442, P. R. China
E-mail: zhoucj@scut.edu.cn

J. Zhu
School of Electronic and Computer Engineering
Peking University
Shenzhen 518055, China

M. Zhang
School of Science and Engineering
The Chinese University of Hong Kong
Shenzhen 518172, China

Cary Y. Yang
Center for Nanostructures
Santa Clara University,
Santa Clara, CA 95053, USA

Self-driven photodetectors (SDPDs) transform photon energy into electrical energy without external voltage, which makes them highly advantageous for applications such as low-power communication and imaging systems. Two-dimensional materials (2DMs) provide ideal platforms for SDPDs thanks to their band structures covering ultraviolet to infrared spectrum, strong light absorption efficiencies, and high carrier mobilities. However, the lack of stable doping methods and the complicated 2DMs multilayer stacking techniques pose tremendous difficulties for 2DMs to adopt the same device structures (i.e. PN junctions) as bulk materials, and the resultant self-driven performance remains at a low level. This work reveals how different asymmetrical effects can be combined to synergistically boost self-driven properties based on typical 2D metal-semiconductor-metal (MSM) photodetectors. Using $WSe_2$ as an exemplary 2D material to build MSM photodetectors, the synergistic effect of asymmetrical contact electrodes and asymmetrical contact geometries is theoretically and experimentally demonstrated. The open-circuit voltage ($V_{oc}$) of the SDPD reaches 0.58V, with a zero-bias responsivity of 5.77 A/W and an on/off ratio of $1.73\times10^5$. Additionally, our devices demonstrate potential for visible light communication (VLC) in underwater environments. Our results offer a promising and efficient strategy for building SDPDs based on various 2DMs and pave the way toward low-power optoelectronic applications.

## 1. Introduction

Self-driven photodetectors (SDPDs) have wide applications in photoelectric detection, image processing, and light communication.[1–5] Traditional bulk semiconductor-based[6–9] manufacturing platforms for photodetectors suffer from limitations such as weak light absorption ability, limited detection spectrum, and long response time. With the increasing performance requirements of PDs, two-dimensional materials (2DMs) have gradually entered the spotlight. They possess excellent carrier mobility, superior optical absorption coefficients,[10] and reduced thicknesses compared to traditional bulk materials. Moreover, 2DMs provide a rich library of band structures for PDs targeting different wavelengths.[11] These advantages make 2DMs-SDPDs promising candidates for next-generation optoelectronic devices.

Recently, research on 2DMs-SDPDs has made some significant progress.[12] The most common approach to introducing self-driven characteristics is to construct PN homojunctions,[13–15] which usually require doping. The mature ion-implantation doping technology cannot be readily applied to 2DMs as it will induce substantial defects[16,17] for ultrathin layered structures. Heterojunctions[18–20] are also applied to realize self-driven properties, which are commonly fabricated by time-consuming 2DMs stacking techniques. The epitaxy growth technology is an alternative method for the construction of 2DMs-heterojunctions, but it can only be achieved under stringent conditions.[21] These methods are generally complex and have poor reproducibility. While photoelectrochemical (PEC) PDs have simpler fabrication processes than others, they only operate in liquid environments.[12] Coupling the optical field with other physical fields can also achieve zero-bias detection, such as piezo-photoelectric detectors, photothermal detectors, flexoelectric photodetectors, and ferroelectric photodetectors.[22–25] These detectors require additional energy input, which increases power consumption and fails to realize true self-driven characteristics. To overcome these shortcomings, several PDs with asymmetrical structures[26] based on metal-semiconductor-metal (MSM) structures have been proposed. Specifically, asymmetrical Schottky barriers,[27,28] contact engineering,[29] and thicknesses[30] lead respectively to self-driven characteristics. The self-driven performance and responsivity of such devices with single asymmetry remain weak, and it is an open question whether these different self-driven mechanisms can be combined synergistically to boost self-driven properties while still preserving the advantages such as simple fabrication, reproducibility, and low-power input.

In this work, the synergistic asymmetrical effect is first proposed to realize 2DMs-SDPDs. The theoretical framework is built and verified with extensive experimental data. We combine different electrodes (Pt and Cr) with asymmetrical contact lengths to design MSM PDs based on a typical 2DM $WSe_2$. The device fabrication process is simple and highly reproducible, requiring no doping or extra physical fields during the operation. Our device can achieve an open-circuit voltage ($V_{oc}$) of 0.58 V and an on/off ratio of $1.73 \times 10^5$. At zero bias, the responsivity (R) and detectivity (D*) reach 5.77 A/W and $1.87 \times 10^{11}$ Jones, respectively. We demonstrate the device's application for underwater coding, highlighting its potential for visible light communication (VLC) in liquid environments. Our research provides a prototype for the design of photodetectors with superior self-driven performance and promising applications.

## 2. Results and Discussion

### 2.1. Theoretical analysis of synergistic asymmetrical effects

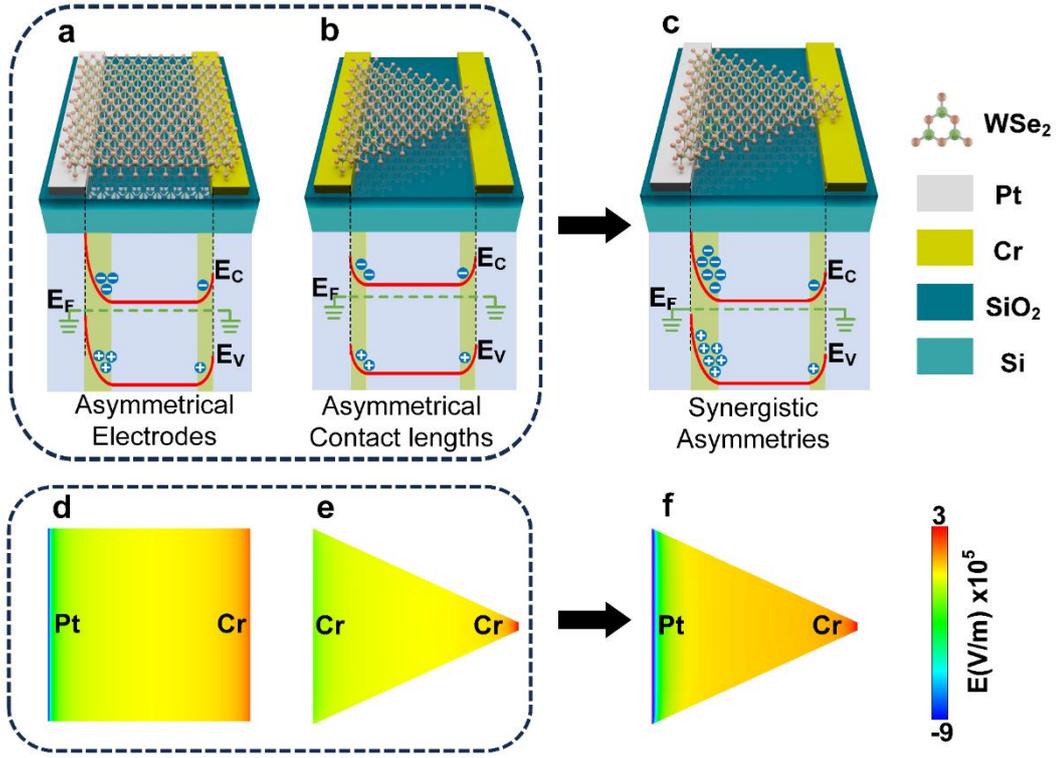

**Figure 1.** Schematics of PDs with different types of asymmetries. a) PDs with asymmetrical electrodes. b) PDs with asymmetrical contact lengths. c) PDs with synergistic asymmetrical effects. d-f) Simulations of electric field distribution corresponding to a-c).

Our MSM PDs feature two bottom electrodes with WSe$_2$ positioned on top. This WSe$_2$-above-electrode structure alleviates the Fermi Pinning effect.[31,32] **Figure 1a-c** compare three different self-driven MSM structures and elucidate the synergistic asymmetrical mechanism of Pt-WSe$_2$-Cr SDPDs. As for self-driven properties, both short-circuit current ($I_{sc}$) and open-circuit voltage ($V_{oc}$) are performance indicators. $I_{sc}$ is thickness-dependent, as shown in Supplementary Note 1. When it comes to the performance comparison of multiple devices, we have to choose the devices with the same thicknesses. In comparison, thickness (t) is not included in the formula related to $V_{oc}$, thus $V_{oc}$ is a more suitable indicator for comparing multiple devices. The derived expression of $V_{oc}$ considering both asymmetrical effects is given by (Supplementary Note 1):

$$V_{oc} = \frac{kT}{q}\ln\left(\frac{I_{ph\_Pt}-I_{ph\_Cr}}{-I_{s\_Pt}+I_{s\_Cr}}\right) = \frac{kT}{q}\ln\left(\frac{qG}{A^*T^2}\frac{L_{Pt}\sqrt{\frac{2\epsilon_s(V_{BiPt})}{qN}}-L_{Cr}\sqrt{\frac{2\epsilon_s(V_{BiCr})}{qN}}}{-L_{Pt}e^{\frac{-q\phi_{Pt}}{kT}}+L_{Cr}e^{\frac{-q\phi_{Cr}}{kT}}}\right) \quad (1)$$

Here, $\frac{kT}{q}$ is the thermal voltage, $I_{ph\_Pt}$, $I_{ph\_Cr}$ are the photocurrents from the Pt and Cr sides, respectively, and $I_{s\_Pt}$, $I_{s\_Cr}$ are the reverse saturation currents, respectively. Further, $G$ is the carrier generation rate, $A^*$ is the effective Richardson constant, $L$ is the contact length of the Schottky junction interface, $\epsilon_s$ is the relative dielectric constant, $N$ is the doping concentration, $V_{Bi}$ is the built-in potential, and $\emptyset$ is the Schottky barrier height (SBH). Asymmetrical Schottky barriers and contact lengths lead to self-driven properties separately, and synergistic asymmetrical effects can be realized when these asymmetries are properly combined. As shown in Figure 1d-f, larger built-in electric field from Pt-WSe$_2$ interfaces are induced by higher Schottky barriers, resulting in higher photocurrent. Then we keep the asymmetrical electrodes (Pt and Cr) unchanged and analyze the influence brought by the contact length differences. The contact length difference $\Delta L$ can be expressed as: $\Delta L = L_{Pt} - L_{Cr}$. When $\Delta L$ is increased, we can extract that the total photocurrent (numerator) increases and the total reverse saturation current (denominator) decreases corresponding to Equation (1), boosting the self-driven performance compared to the single-asymmetry structures. As a result, synergistic asymmetrical effects can be realized when the longer contact lengths are set on the sides with higher Schottky barriers. In the following sections, experimental evidence is provided to verify the validity of the proposed synergistic asymmetrical effects.

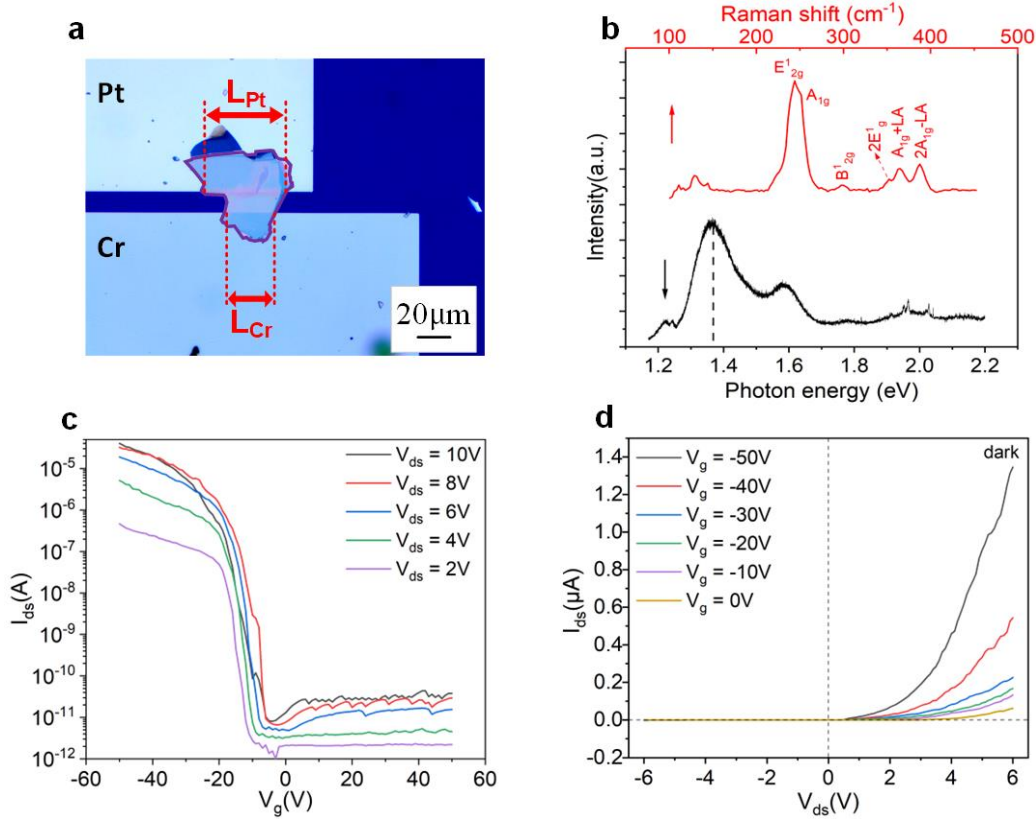

**Figure 2.** Basic characterizations and electrical measurements. a) Optical microscope image of one Pt-WSe$_2$-Cr PD. b) Raman and PL spectra of WSe$_2$. c) Transfer curves in darkness. d) Output curves in darkness.

## 2.2. Electrical measurements of 2D WSe$_2$ photodetectors

**Figure 2a** shows an optical microscope image of a Pt-WSe$_2$-Cr PD. $L_{Pt}$ and $L_{Cr}$ are 47.013 μm and 28.165 μm, respectively, with a contact length difference of 18.848 μm. We have also prepared many other contrast samples, as shown in Figure S1. Figure 2b displays the Raman and PL spectra of WSe$_2$. The E$^1_{2g}$ and A$_{1g}$ peaks are located at 244.0 cm$^{-1}$ and 251.6 cm$^{-1}$, respectively. A high A$_{1g}$ peak is a typical characteristic of multilayer WSe$_2$.[33] Other peaks, 2E$^1_g$, A$_{1g}$+LA, and 2A$_{1g}$-LA are located at 352.5 cm$^{-1}$, 365.1 cm$^{-1}$, and 387.6 cm$^{-1}$, respectively, which are consistent with previous reports.[34] Thickness-dependent bandgap has been reported in WSe$_2$.[35] With the increase of the thickness, WSe$_2$ is transformed into an indirect bandgap semiconductor. There are two peaks in our as-fabricated material, indicating indirect transition (IT) peak and direct transition (DT) peak, 1.36eV and 1.6eV, respectively, which is consistent with other reported works. [29,36] For many semiconductor materials, the main peak in PL spectrum usually corresponds to near-band-edge emission due to the most efficient absorption.[37] Therefore, the bandgap of the multi-layer WSe$_2$ is equal to 1.36eV. The film thickness is approximately 41 nm (Figure S2). First, several three-terminal measurements are conducted. Transfer curves under dark conditions are plotted in Figure 2c. Due to Schottky barriers on both sides, this device exhibits low dark current (I$_{dark}$) and a high on-off ratio (~10$^6$).[28] Although many reports on WSe$_2$-based transistors exhibited bipolar transport properties, electron transport is suppressed in our devices. This is mainly due to the MS contact interface states and the metals we choose.[31] The Cr electrode has a significant lattice mismatch with most transition metal dichalcogenides (TMDs) in general.[38] Some related works also reported that the Cr-WSe$_2$ MS contact Schottky junctions only exhibited hole transport properties.[39] Previous calculations have suggested different Schottky barrier heights for electrons and holes in metal-WSe$_2$ contacts. For Pt-WSe$_2$ contacts, hole Schottky barriers are much smaller than electron Schottky barriers,[40] which results in predominantly hole transport across this interface. Therefore, we believe the p-type transport transfer curve of the Pt-WSe$_2$-Cr MSM PDs aligns with prior experiments and theoretical calculations. Output curves are then measured under dark and illuminated conditions, as shown in Figure 2d and Figure S3a. The gate terminal maintains carrier modulation capability under both circumstances. Upon illumination, V$_{oc}$ reaches about 0.40 V due to photo-generated carriers. To better visualize the gate-tunable photoresponse property, I$_{ds}$ at V$_{ds}$ = 6V under darkness and illumination are extracted and plotted in Figure S3b. As the gate voltage increases from -10 V to -50 V, I$_{ph}$ rises from 0.36 µA to 3.78 µA and the difference between I$_{ph}$ and I$_{dark}$ also increases. In the following section, we use PDs as two-terminal devices and focus on their self-driven performance.

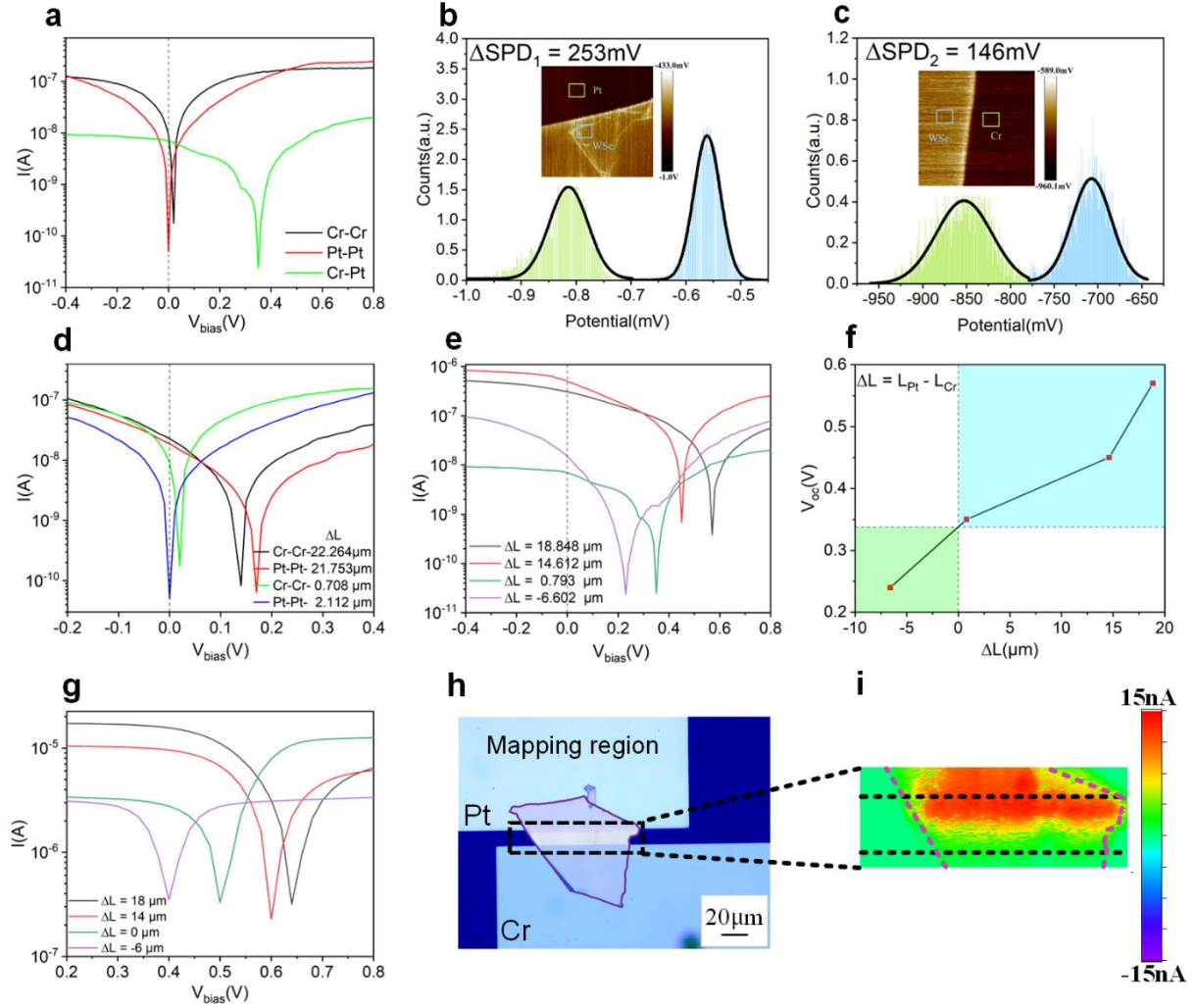

**Figure 3.** The verifications of synergistic asymmetrical effects. a) I-V curves of SDPDs with the same and different electrodes. b-c) Gaussian fitting curves of the surface potential differences. Insets show corresponding KPFM results of Pt-WSe$_2$ and WSe$_2$-Cr interfaces. d) I-V curves of SDPDs with the same electrodes and different contact lengths. e) I-V curves of Pt-WSe$_2$-Cr SDPDs with different contact lengths. f) The relationship between V$_{oc}$ and Δ$L$ of Pt-WSe$_2$-Cr SDPDs. g) Simulated I-V curves of Pt-WSe$_2$-Cr SDPDs with different contact lengths. h) Optical image of the Pt-WSe$_2$-Cr SDPD. i) Photocurrent mapping image corresponding to black dash square in h at zero bias.

## 2.3. Experimental verifications of the synergistic asymmetrical effect

For the two-terminal photocurrent I-V measurements, we use a 532 nm laser with fixed power intensity as the light source. The size of the light spot is large enough to cover both Schottky barriers. To maintain consistency in discussions, Cr side is connected to the ground. For PDs with the same electrodes but different contact lengths, Δ$L$ is the difference between the longer and the shorter electrode. The entire device is denoted as Metal$_1$-Metal$_2$-Δ$L$. For PDs with Pt-Cr electrodes, the V$_{oc}$ is approximately 0.35 V, while PDs with the same electrodes have very weak self-driven characteristics, as shown in **Figure 3a**. This is a result of the asymmetrical Schottky barriers across the two electrode interfaces. We then quantitatively evaluate the surface potential differences of Pt-WSe$_2$ and WSe$_2$-Cr interfaces, respectively, as shown in Figure 3b-c. The workfunction difference between Pt and WSe$_2$ is 253 mV ($\Delta SPD_1$), and that between Cr and WSe$_2$ is 146 mV ($\Delta SPD_2$). The Schottky barrier difference ($\Delta$SBH) between two sides can be expressed as $\Delta SBH = \Delta SPD_1 - \Delta SPD_2$ (Supplementary Note 2). Thus, $\Delta SBH = 107$ mV, which is one factor leading to self-driven properties, despite the deviation from the Schottky-Mott rule. In addition, different contact lengths also result in self-driven characteristics. Due to asymmetrical effective contact areas of MS junctions, the Cr-Cr-22.264μm and Pt-Pt-21.753μm devices have V$_{oc}$ of 0.14 V and 0.17 V, respectively, while PDs with symmetrical contact lengths have a maximum V$_{oc}$ of only 0.02 V, as shown in Figure 3d. There is a debate on the working mechanism of geometrically asymmetric devices. To figure out the predominant factor that leads to the self-driven performance, the KPFM characterizations of devices with the same electrodes but different contact lengths are conducted, as shown in Figure S4. For device Cr-Cr-22.264 μm in Figure S4a-c, the surface potential differences on two sides are 139mV and 146mV, respectively. For device Pt-Pt-21.753 μm in Figure S4d-f, the surface potential differences on two sides are 239mV and 251mV, respectively. For device Pt-Cr-14.612 μm in Figure 3b, c, the surface potential differences on two sides are 253mV and 146mV, respectively. Due to the fermi pinning effect, the measured surface potential differences are smaller than ideal Schottky heights.[32,41,42] The surface potential differences exhibit slightly asymmetric in these two devices. This is due to negligible asymmetrical distribution of surface states.[43,44] Compared to other reported papers, we believe that ~10mV cannot be the predominant factor that leads to the self-driven performance in our devices.[32,41,42] Therefore, we believe that the self-driven performance of these devices is mainly attributed to the asymmetrical contact lengths.[29,45] To verify the feasibility of synergistic asymmetrical effects, a series of PDs with both asymmetrical Schottky barriers and asymmetrical contact lengths are then fabricated. As for PDs with different electrodes, $\Delta L$ is defined as the difference between Pt-WSe$_2$ side and WSe$_2$-Cr side, denoted by $\Delta L = L_{Pt} - L_{Cr}$. As shown in Figure 3e, those with a positive $\Delta L$ have a V$_{oc}$ greater than 0.35 V, exhibiting synergistic asymmetrical effects. For the Pt-Cr-18.848μm device, V$_{oc}$ reaches approximately 0.58 V. In contrast, the self-driven performance of the device Pt-Cr-(-6.602μm) is much lower. Due to the larger contact length from the Cr interface, $I_{ph\_Cr}$ and $I_{s\_Cr}$ across the WSe$_2$-Cr interface increase, resulting in decrease in the total photocurrent and increase in the total reverse saturation current, thus reducing the self-driven capabilities. To further demonstrate the validity, the dark currents and their rectification ratios of these samples are also measured, as shown in Figure S5 and Table S1. Devices with asymmetrical Schottky barriers, whether different

Schottky barrier height or asymmetrical contact lengths, exhibit rectification effect in dark current measurements. Compared to samples with only asymmetrical contact lengths (Figure S5a, c), samples with synergistic asymmetrical effect (Figure S5e, f, g and Figure 2a) exhibit larger rectification ratio, as shown in Table S1. Larger rectification ratio means better self-driven performance under illuminated environment, which is consistent with the theory. To visualize the self-driven performance brought by double asymmetries, the $V_{oc}$-$\Delta L$ curve is shown in Figure 3f. Finite element method (FEM) is then used to simulate the self-driven performance of PDs with different degrees of asymmetry, as shown in Figure 3g. The variation tendency of self-driven performance is related to the polarity of $\Delta L$, showing similar trends with the experimental results. Other simulation results are summarized in Figure S6a-b. Further, we conduct photocurrent mapping at zero bias, as shown in Figure 3i and Figure S7. For PDs with different electrodes, photocurrents from Pt-$WSe_2$ interfaces dominate, while Cr-$WSe_2$ interfaces exert negligible influence. Photocurrents with the same electrodes are relatively balanced on the two sides. These results help visualize the contributions of self-driven performance from two different MS interfaces and further verify the validity of synergistic asymmetrical effects.

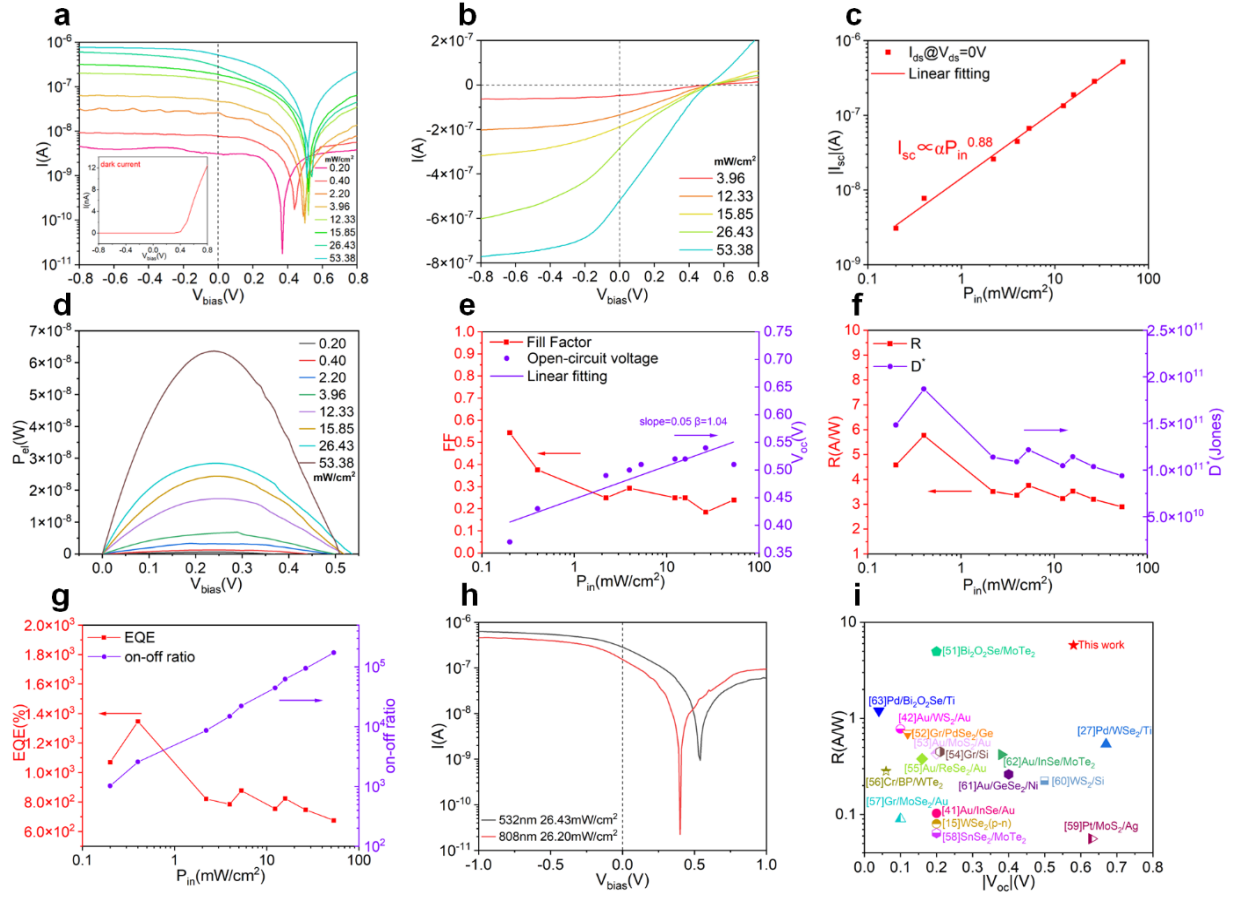

**Figure 4.** The photoresponse measurements of the Pt-WSe$_2$-Cr SDPDs. a) Semi-logarithmic I-V curves under various light power densities. The inset shows the dark I-V curve. b) Linear plot of I-V curves under varied light power intensities. c) Relationship between $I_{sc}$ and light power intensity $P_{in}$. d) $P_{el}$-V curves under varied light power intensities. e) Fill factor and $V_{oc}$ with varying $P_{in}$ and linear fit of $V_{oc}$-$P_{in}$. f) R and D* with the variation of $P_{in}$. g) EQE and on-off ratio with the variation of $P_{in}$. h) Semi-logarithmic I-V curves under different wavelength lights. i) Comparison of performances between our work and other reports. [15,27,41,42,51-63]

In addition, we systematically measure the photoresponse of SDPDs with synergistic asymmetrical effects. Semi-logarithmic and linear I-$V_{bias}$ curves under different light power intensities are plotted in **Figure 4a-b**, respectively. Self-driven performance is enhanced with the increase of the light power intensity. As the power intensity increases from 0.20 mW/cm$^2$ to 53.38 mW/cm$^2$, $I_{sc}$ rises from 3.05 nA to 519.5 nA. The absolute values of $I_{sc}$ under different light power densities are extracted and plotted in Figure 4c. We also conduct linear fitting using the equation $I = \alpha P_{in}^{\theta}$,[19] where $\theta$ = 0.88, indicating sub-linear properties. This is related to the number of defect centers in the channel and the efficiency of photogenerated carrier separation. Additionally, lateral devices exhibit lower photo-gain compared to vertical structures, further degrading the linearity. Since our devices demonstrate the photovoltaic (PV) effect, the output electrical power $P_{el}$-$V_{bias}$ curves under varied power intensities are plotted in

Figure 4d. $P_{el}$, expressed as $P_{el} = IV_{bias}$,[15] reaches the maximum value of 63.6 nW when $V_{bias}$=0.26 V. Fill factor (FF) is also a crucial parameter to evaluate the output performance for devices with PV effect,[14] defined as $FF = \frac{\max(P_{el})}{I_{sc}V_{oc}}$, where max $(P_{el})$ is the maximum value of the $P_{el}$. Typically, PDs exhibit higher output efficiency under weak light intensity. FF reaches its highest at 0.20 mW/cm², with a value of 0.54. The linear fitting of $V_{oc}$ is plotted on the right side of Figure 4e. The slope can be given by $\frac{dV_{oc}}{d\ln P} = \frac{2}{\beta}\frac{k_0 T}{q}$, where the fitted value of $\beta$ indicates different recombination mechanisms.[46] When $\beta$ approaches 1, Shockley-Read-Hall (SRH) recombination is predominant. This usually happens in narrow-bandgap semiconductors. When $\beta$ approaches 2, the Langevin recombination model becomes dominant. A value of $\beta$ greater than 2 demonstrates that defect-assisted recombination cannot be ignored. Our device fits with $\beta = 1.04$, which indicates an SRH-dominant mechanism.

We then analyze and calculate other key parameters, as shown in Figure 4f-g. Responsivity represents the device's detection efficiency,[6] defined as $R = \frac{I_{ph}}{P_{in} \times S}$, where the photocurrent $I_{ph}$ is the difference between total current and $I_{dark}$, and S is the effective illuminated area of the device (336 μm², Fig. 2a). Detectivity measures the ability of weak light detection,[6] given by $D^* = R\sqrt{\frac{S}{2qI_{dark}}}$, which is proportional to $R$. At zero bias, $R$ and $D^*$ reach the maximum values of 5.77 A/W and 1.87×10¹¹ Jones under 0.40 mW/cm² illumination, respectively. As the power intensity increases, $R$ and $D^*$ exhibit downward trends. This is due to the saturation of photogenerated carriers under strong power intensity, which reduces carrier separation rate. Defect energy levels in the material can also affect carrier separation efficiency.[47] External quantum efficiency (EQE) is the ratio of electron-hole pairs generated in the channel per unit time to the number of photons emitted by the light source. EQE indicates photoelectric conversion efficiency,[6] given by $EQE = \frac{Rhc}{q\lambda}$, where $R$ is the responsivity, h is Planck's constant, c is the speed of light in vacuum, and $\lambda$ is the incident light wavelength. The EQE of our device is greater than 100% under different power levels, which indicates that one photon can create more than one electron-hole pair. EQE reaches a maximum of approximately 1400% at 0.40 mW/cm². High photocurrent amplification (gain) can be achieved when there are deep traps on the surface of channel materials, which leads to long carrier recombination lifetimes.[6,48] Given that relatively slow response time (~30ms, Figure S9d) of our devices compared to others,[28,41,49,50] we believe the real charge recombination lifetime is significantly larger than the ideal lifetime, which gives an explanation for the boosting EQE. At zero bias, the on-off ratio reaches 1.73 ×10$^5$ at 53.38 mW/cm$^2$, which exhibits desirable switching characteristics. Subsequently, we compare the photoresponse between visible (532 nm) and near-infrared (808 nm) light. The photoresponse of 532 nm light is stronger than that of 808 nm under basically the same power intensity, consistent with our simulation results (Figure S3c). Compared to other reported photodetectors[15,27,41,42,51-63] (Figure 4i), our devices exhibit superior self-driven performance and desirable responsivity. In addition to photoresponse, the time characteristics are also critical to PDs. Bandwidth is an important performance indicator for photodetectors. For most photodetectors, the responsivity decays as the frequency of the incident light increases. Bandwidth is defined as the corresponding frequency when the responsivity drops to 0.707. [6] The bandwidth of our device is 170Hz, as shown in Figure S8. We measure the switching stability of the device using a 532 nm laser with a duty cycle of 50% and a frequency of 0.5 Hz, as shown in Figure S9a-b. This device exhibits excellent stability over 500 seconds. The normalized photoresponse under different power intensities is illustrated in Figure S9c. There, the response time is measured at a frequency of 1 Hz without external bias. The rise and fall times are 25.8 ms and 34.8 ms, respectively, as shown in Figure S9d.

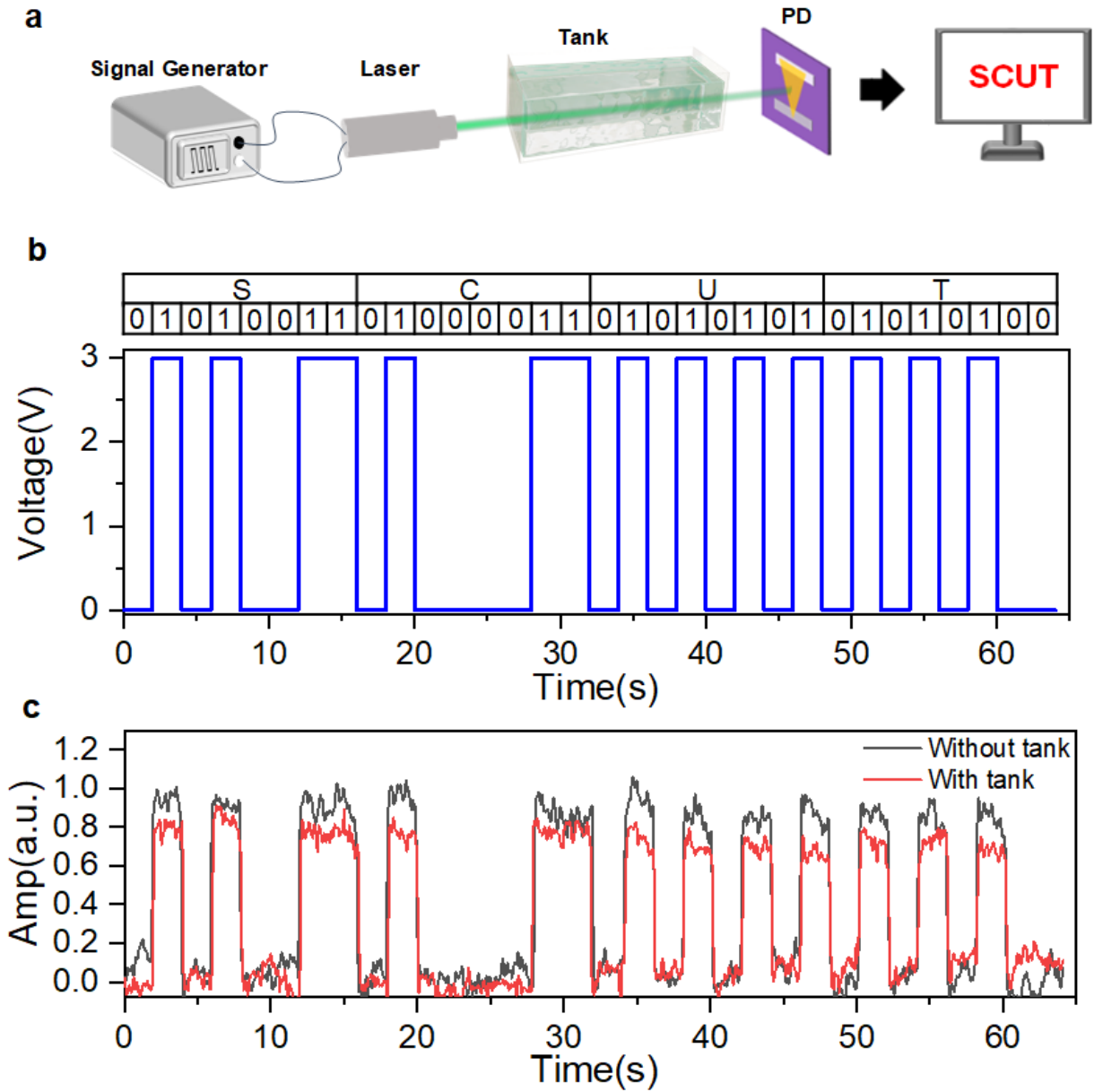

**Figure 5.** Underwater visible light communication based on Pt-WSe$_2$-Cr SDPDs. a) Schematic diagram of the optical path for simulating underwater visible light communication. b) The ASCII bitstream of "SCUT" encoded by the signal generator. c) Normalized photoresponse of the SDPD in ambient and underwater conditions.

## 2.4. Underwater communication based on Pt-WSe$_2$-Cr photodetectors

Owing to simple fabrication processes and enhanced self-driven performance, our devices show potential in light communication in certain passive environments. While near-infrared (NIR) band is widely used, visible light communication (VLC) has its advantages. Compared to NIR spectrum, the visible (VIS) spectrum is abundant and does not require commercial licensing. In underwater environments, NIR light suffers from severe attenuation, which limits its transmission distance. On the contrary, green and blue light have transparent windows underwater, which allows them to propagate over long distances. The bandgap of multi-layered

WSe$_2$ also fits with the VIS spectrum. To show the feasibility of underwater VLC of our SDPDs, an experimental platform is shown in **Figure 5a**. The ASCII bitstream of "SCUT" is programmed by a signal generator. Then the bitstream is sent to our PD through this system. Finally, the ASCII bitstream is decoded into the alphabets "SCUT". At zero bias, normalized photoresponse is measured with and without a water tank in-between, respectively, as shown in Figure 5c. Despite a slight decrease in photoresponse, our device shows desirable performance in underwater VLC without an external power supply.

## 3. Conclusion

In summary, we have developed self-driven Pt-WSe$_2$-Cr photodetectors with synergistic asymmetrical effects. By properly combining asymmetrical electrodes with asymmetrical contact engineering, enhanced self-driven performance is achieved. Our device has a V$_{oc}$ of 0.58V. A responsivity of 5.77A/W and a detectivity of 1.87×10$^{11}$ Jones are measured respectively under zero bias. The fabrication process without any special requirements of doping and heterostructures offers an effective method for designing SDPDs based on various 2DMs. The concept of synergistic asymmetrical effect provides a universal method for designing photodetectors with enhanced self-driven performance. Furthermore, our devices show potential in underwater VLC and other passive application scenarios. These desirable properties make our SDPDs highly competitive among low-power optoelectronic devices.

## 4. Experimental Section

**4.1. Fabrication of MSM photodetectors.** A 4-inch clean Si/SiO$_2$ (300 nm) wafer is prepared and treated with plasma. After that, two rounds of standard photolithography are performed to deposit asymmetrical electrodes, Cr(50nm) and Pt(50nm), respectively. Then multilayer WSe$_2$ flakes (30-100nm) are obtained by mechanical exfoliation and transferred onto electrodes through PDMS-assisted dry transfer method. MSM photodetectors can be fabricated with steps mentioned before.

**4.2. Material characterizations and optoelectronic measurements.** We use an optical microscope (Axio Scope A1, Zeiss) with x50 objective lens to capture schematics of device structures. Horiba Spectrophotometer system is used to obtain Raman and PL spectra of WSe$_2$. The thicknesses and surface potential differences of WSe$_2$-Cr and WSe$_2$-Pt interfaces are characterized by Atomic Force Microscopy (AFM) and frequency modulation Kelvin Probe Force Microscopy (FM-KPFM) (Bruker Dimension Icon). Optoelectronic responses are measured through the optoelectronic platform with optoelectronic system software packages (TuoTuo Technology (Suzhou) Co. Ltd). For the photocurrent mappings, the step size of x and y direction is 1.6μm and 1.2μm, respectively. Light power intensity is measured by a digital power meter (Thorlabs PM 100D). Electrical measurements are conducted by a semiconductor analyzer (Keithley 4200A-SCS). A signal generator (Tektronix AFG 31252) is employed to generate programmed bitstreams. A lock-in amplifier (OE 1201) and an oscilloscope (PicoScope 5244D) are used to collect the photoresponse at zero bias. For switching stability tests, the frequency of low-pass filter is set at 60 Hz. For the time response measurement, the frequency of low-pass filter is set at 10Hz. All the characterizations and measurements are conducted at room temperature.

**Supporting Information**

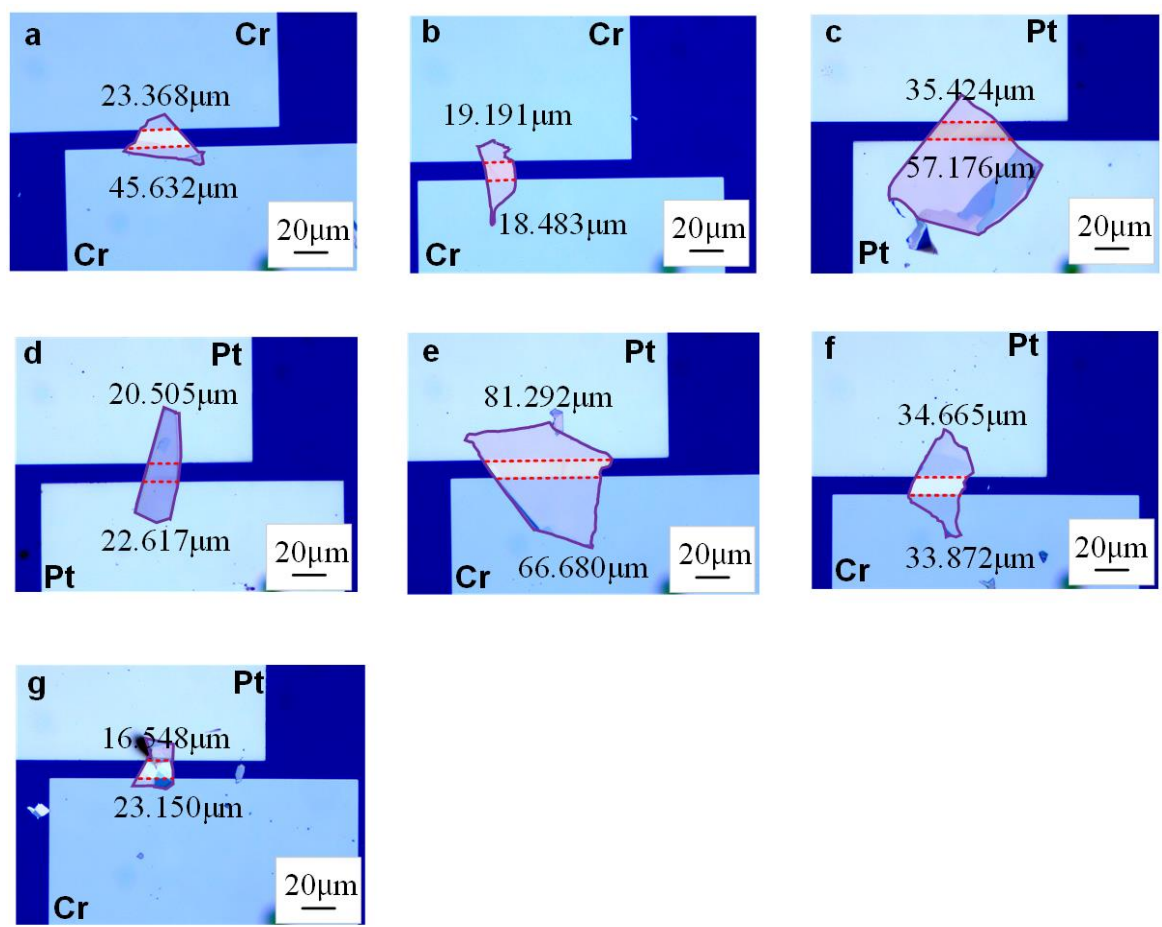

**Figure S1.** Optical images of a series of MSM WSe$_2$ photodetectors. a) Cr-Cr-22.264μm. b) Cr-Cr-0.708μm. c) Pt-Pt-21.752μm. d) Pt-Pt-2.112μm. e) Pt-Cr-14.612μm. f) Pt-Cr-0.793μm. g) Pt-Cr- (-6.602)μm.

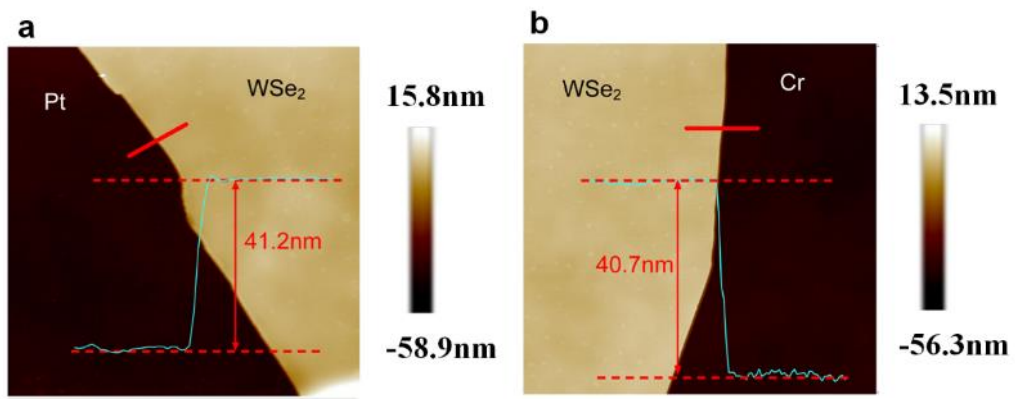

**Figure S2.** Thickness measurements of WSe$_2$-metal interfaces. a) The region of the Pt-WSe$_2$ interface. b) The region of the WSe$_2$-Cr interface.

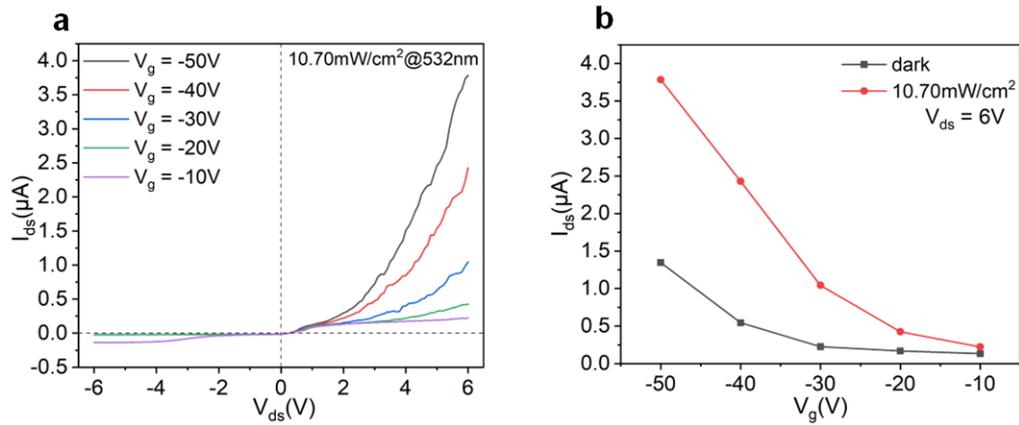

**Figure S3.** Photoresponse under three-terminal measurement. a) Output curve at light power intensity of 10.70 mW/cm². b) $I_{ds}$-$V_g$ curve at $V_{ds}$ = 6V in darkness and at a light power intensity of 10.70 mW/cm².

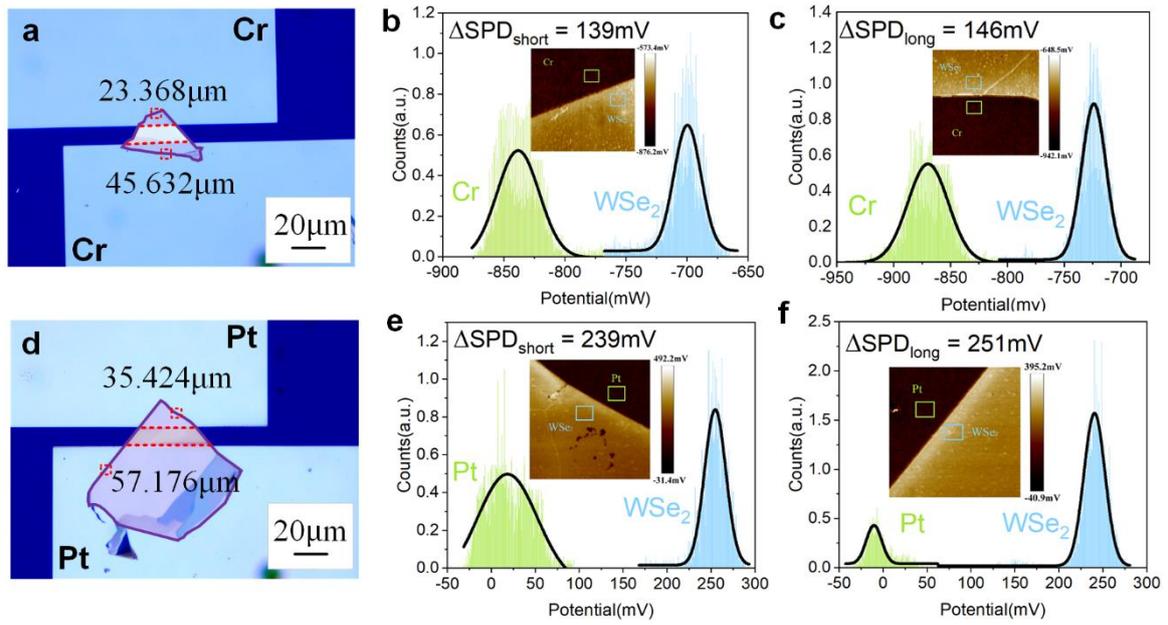

**Figure S4.** FPFM characterization of devices with the same electrodes and different contact lengths. a) Cr-Cr-22.264μm. b-c) Gaussian fitting curves of the surface potential differences in a). Insets show corresponding KPFM results of upper and lower interfaces. d) Pt-Pt-21.752μm. e-f) Gaussian fitting curves of the surface potential differences in d). Insets show corresponding KPFM results of upper and lower interfaces.

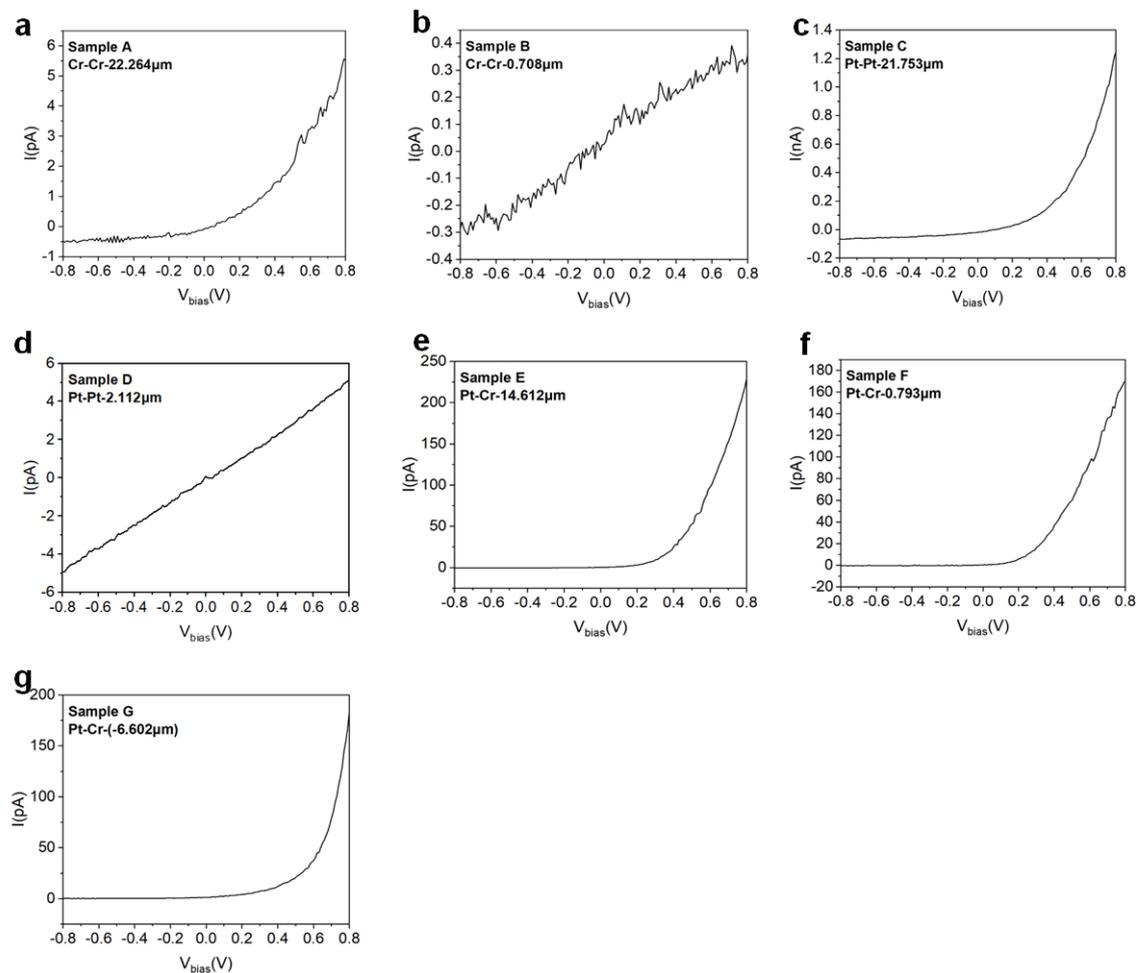

**Figure S5.** Dark current measurement of samples corresponding to Figure S1.

| Sample | Rectification ratio | Sample | Rectification ratio |
|---|---|---|---|
| A | 11.44 | B | - |
| C | 18.65 | D | - |
| E | 590.2 | F | 467.85 |
| G | 395.65 | Figure 2a | 1007.3 |

**Table S1. D**ark current rectification ratio corresponding to Figure S5.

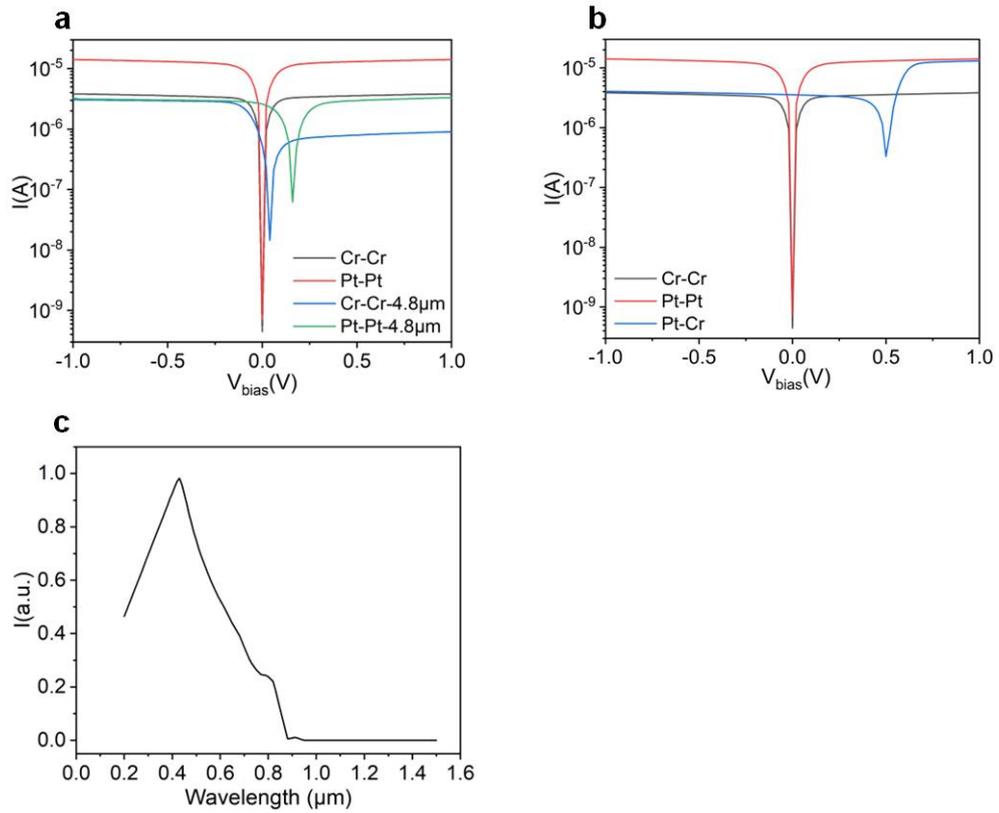

**Figure S6.** Simulation results based on the finite element method. a) Self-driven performance comparison of the same and different contact lengths. b) Self-driven performance comparison of the same and different electrodes. c) The relationship between normalized photoresponse and the wavelength of incident light.

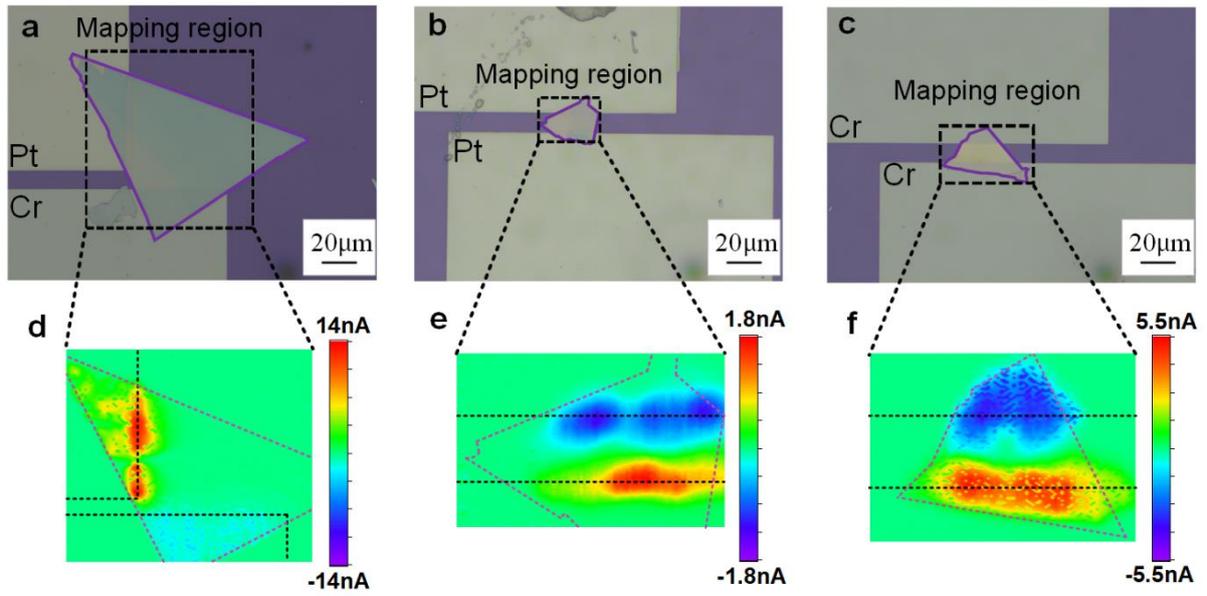

**Figure S7.** Schematic diagrams of three MSM WSe$_2$ photodetectors and the heatmaps of photocurrents at zero bias. a) Pt-Cr, b) Pt-Pt, c) Cr-Cr are three photodetectors with different electrode combinations. d-f) are photocurrent mapping images which correspond to a-c).

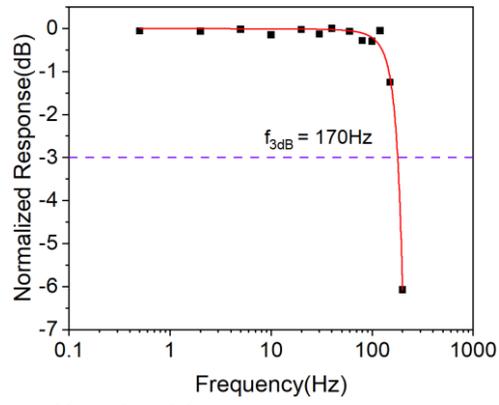

**Figure S8.** The measurement of bandwidth.

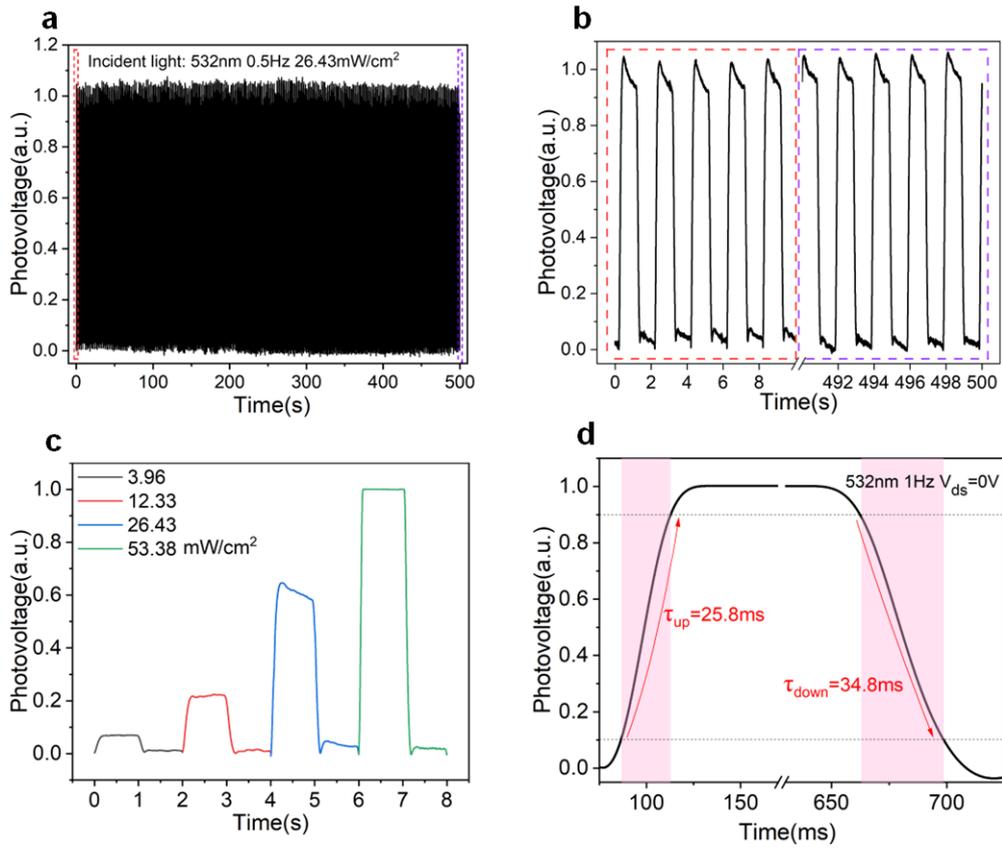

**Figure S9.** Time response measurements of MSM SDPDs. a) The switching stability test. b) The switching stability of the first and the last five cycles. c) Normalized photoresponse under varied power densities. d) The response time of the PD.

## Supplementary Note 1

When $I_{sc}$ is used as the indicator to measure the self-driven performance, the formula can be given by (assuming the photocurrent from Pt-WSe$_2$ side is positive):

$$I_{sc} = (J_{sc\_Pt} \times L_{Pt} - J_{sc\_Cr} \times L_{Cr}) \times t \tag{1}$$

, where $J_{sc\_Pt}$ and $J_{sc\_Cr}$ are the current densities of the Pt-WSe$_2$ and WSe$_2$-Cr interfaces. $L_{Pt}$ and $L_{Cr}$ are the contact lengths of two MS interfaces, and $t$ is the thickness of the WSe$_2$ flake. It can be concluded that $I_{sc}$ is a thickness-dependent performance indicator.

To avoid this problem, theoretical derivations are based on $V_{oc}$, which is calculated as follow:

$$V_{oc} = \frac{kT}{q} \ln\left(\frac{I_{ph}}{I_s}\right) \tag{2}$$

, where $\frac{kT}{q}$ is the thermal voltage. $I_s$ and $I_{ph}$ are total photocurrent and inverse saturation current, respectively. We first analyze a metal-semiconductor (MS) Schottky barrier on one side. $I_s$ can be given by:

$$I_s = LtA^*T^2 \exp\left(-\frac{\emptyset}{kT}\right) \tag{3}$$

, where $L$ is the contact length of Schottky junction interface, $t$ is the thickness of WSe$_2$, $A^*$ is the effective Richardson constant, $\emptyset$ is the Schottky barrier height. The dark current of one WSe$_2$-metal interface can be expressed as:

$$I_{dark} = I_s \left[\exp\left(\frac{qV}{kT}\right) - 1\right] \tag{4}$$

The photocurrent can be expressed as:

$$I_{ph} = wtqG \sqrt{\frac{2\epsilon_s(V+V_{bi})}{qN}} \tag{5}$$

, where $q$ is the element charge, G is the carrier generation rate, $\epsilon_s$ is the relative dielectric constant, $N$ is the doping concentration, and $V_{bi}$ is the built-in potential, which can be calculated as follow:

$$V_{bi} = \emptyset - \varphi_n \tag{6}$$

, where $\varphi_n$ is the difference between the Fermi level and the electron affinity of WSe$_2$. Considering the entire MSM structure, $V_{oc}$ can be given by (assuming the photocurrent of Pt-WSe$_2$ side is larger and the inverse saturation current of Pt-WSe$_2$ side is smaller):

$$V_{oc} = \frac{kT}{q} \ln\left(\frac{I_{ph\_Pt}-I_{ph\_Cr}}{-I_{s\_Pt}+I_{s\_Cr}}\right) = \frac{kT}{q} \ln\left(\frac{qG}{A^*T^2} \frac{L_{Pt}\sqrt{\frac{2\epsilon_s(V_{BiPt})}{qN}}-L_{Cr}\sqrt{\frac{2\epsilon_s(V_{BiCr})}{qN}}}{-L_{Pt}e^{\frac{-q\emptyset_{Pt}}{kT}}+L_{Cr}e^{\frac{-q\emptyset_{Cr}}{kT}}}\right) \tag{7}$$

## Supplementary Note 2

To quantitatively measure the Schottky barrier difference on both sides, two interfaces are scanned by Kelvin probe force microscopy (KPFM), as shown in Fig. 3b-c. The probe tip is highly susceptible to interference from ambient water vapor, dust, and electrostatic forces during the measurements. By testing surface potential difference (SPD) instead, we can eliminate these non-ideal factors. The SPD between the probe and sample can be expressed as:

$$SPD_{Sample} = \frac{W_{Probe} - W_{Sample}}{q} \tag{8}$$

, where $W_{Probe}$ and $W_{Sample}$ are the workfunction of the probe and sample respectively, $q$ is the elementary charge. For metal-WSe$_2$ junctions, the SPD on both sides can be expressed as:

$$\Delta SPD_1 = SPD_{Pt} - SPD_{WSe_2} \tag{9}$$

$$\Delta SPD_2 = SPD_{Cr} - SPD_{WSe_2} \tag{10}$$

Here, $SPD_{Pt}, SPD_{Cr}, SPD_{WSe_2}$ represent the workfunction differences between probe and Pt, Cr, WSe$_2$, respectively. $\Delta SPD_1, \Delta SPD_2$ denote the workfunction differences between Pt, Cr and WSe$_2$, respectively. Thus, Schottky barrier difference ($\Delta$SBH) between two sides can be expressed as:

$$\Delta SBH = \Delta SPD_1 - \Delta SPD_2 \tag{11}$$

**References**

[1]  M. Choi, S.-R. Bae, L. Hu, A. T. Hoang, S. Y. Kim, J.-H. Ahn, *Science Advances* **2020**, *6*, DOI 10.1126/sciadv.abb5898.

[2]  T. Li, J. Miao, X. Fu, B. Song, B. Cai, X. Ge, X. Zhou, P. Zhou, X. Wang, D. Jariwala, W. Hu, *Nat. Nanotechnol.* **2023**, *18*, 1303.

[3]  Y. Xia, C. Geng, X. Bi, M. Li, Y. Zhu, Z. Yao, X. Wan, G. Li, Y. Chen, *Advanced Optical Mater.* **2023**, *12*, DOI 10.1002/adom.202301518.

[4]  L. Pi, P. Wang, S.-J. Liang, P. Luo, H. Wang, D. Li, Z. Li, P. Chen, X. Zhou, F. Miao, T. Zhai, *Nat. Electronics* **2022**, *5*, 248.

[5]  J. Zou, Y. Ke, X. Zhou, Y. Huang, W. Du, L. Lin, S. Wei, L. Luo, H. Liu, C. Li, K. Shen, A. Ren, J. Wu, *Advanced Optical Mater.* **2022**, *10*, DOI 10.1002/adom.202200143.

[6]  L. Shi, K. Chen, A. Zhai, G. Li, M. Fan, Y. Hao, F. Zhu, H. Zhang, Y. Cui, *Laser & Photonics Reviews* **2020**, *15*, DOI 10.1002/lpor.202000401.

[7]  M. Billet, Y. Desmet, C. Coinon, G. Ducoumau, X. Wallart, J. F. Lampin, E. Peytavit, in *2017 42nd International Conference on Infrared, Millimeter, and Terahertz Waves (IRMMW-THz)*, **2017**, pp. 1–2.

[8]  S. Lischke, A. Peczek, J. S. Morgan, K. Sun, D. Steckler, Y. Yamamoto, F. Korndörfer, C. Mai, S. Marschmeyer, M. Fraschke, A. Krüger, A. Beling, L. Zimmermann, *Nat. Photonics* **2021**, *16*, 258.

[9]  S. Mauthe, Y. Baumgartner, M. Sousa, Q. Ding, M. D. Rossell, A. Schenk, L. Czornomaz, K. E. Moselund, *Nat. Communications* **2020**, *11*, DOI 10.1038/s41467-020-18374-z.

[10] Z. Cheng, R. Cao, K. Wei, Y. Yao, X. Liu, J. Kang, J. Dong, Z. Shi, H. Zhang, X. Zhang, *Advanced Science* **2021**, *8*, DOI 10.1002/advs.202003834.

[11] M. Long, P. Wang, H. Fang, W. Hu, *Advanced Functional Mater.* **2018**, *29*, DOI 10.1002/adfm.201803807.

[12] H. Qiao, Z. Huang, X. Ren, S. Liu, Y. Zhang, X. Qi, H. Zhang, *Advanced Optical Mater.* **2019**, *8*, DOI 10.1002/adom.201900765.

[13] Y. Kang, Y. Pei, D. He, H. Xu, M. Ma, J. Yan, C. Jiang, W. Li, X. Xiao, *Light: Science & Applications* **2024**, *13*, DOI 10.1038/s41377-024-01477-3.

[14] L. Jian, S. Zhang, W. Gao, Y. Sang, Y. Sun, N. Huo, Z. Zheng, M. Yang, *Appl. Phys. Letters* **2024**, *124*, DOI 10.1063/5.0190442.

[15] Y. Tang, Z. Wang, P. Wang, F. Wu, Y. Wang, Y. Chen, H. Wang, M. Peng, C. Shan, Z. Zhu, S. Qin, W. Hu, *Small* **2019**, *15*, DOI 10.1002/smll.201805545.

[16] H. Yoo, K. Heo, M. H. R. Ansari, S. Cho, *Nanomaterials* **2021**, *11*, 832.

[17] K. Zhang, J. Robinson, *MRS Advances* **2019**, *4*, 2743.

[18] Y. Mu, J. Yang, G. Xie, Z. Wang, B. Guo, J. R. Gong, *Advanced Functional Mater.* **2024**, *34*, DOI 10.1002/adfm.202315543.

[19] C. Zhang, S. Peng, J. Han, C. Li, H. Zhou, H. Yu, J. Gou, C. Chen, Y. Jiang, J. Wang, *Advanced Functional Mater.* **2023**, *33*, DOI 10.1002/adfm.202302466.

[20] K. Xiao, S. Zhang, K. Zhang, L. Zhang, Y. Wen, S. Tian, Y. Xiao, C. Shi, S. Hou, C. Liu, L. Han, J. He, W. Tang, G. Li, L. Wang, X. Chen, *Advanced Science* **2024**, DOI 10.1002/advs.202401716.

[21] Z. Zhang, X. Yang, K. Liu, R. Wang, *Advanced Science* **2022**, *9*, DOI 10.1002/advs.202105201.

[22] D. Q. Zheng, Z. Zhao, R. Huang, J. Nie, L. Li, Y. Zhang, *Nano N.a.* **2017**, *32*, 448.

[23] M. Dai, X. Zhang, Q. J. Wang, *Advanced Functional Mater.* **2024**, *34*, DOI 10.1002/adfm.202312872.

[24] L. Qi, W. Tang, X. Weng, K. Wu, Y. Cen, Y. Sun, S. Zhou, Z. Li, X. Wu, C. Kang, D. Zhao, S. Dai, Y. Xie, H. Liang, W. Zhang, Y. Zeng, S. Ruan, *Advanced Functional Mater.* **2024**, *34*, DOI 10.1002/adfm.202315991.

[25] S. Wang, L. Li, W. Weng, C. Ji, X. Liu, Z. Sun, W. Lin, M. Hong, J. Luo, *J. Of The American Chemical Society* **2019**, *142*, 55.

[26] Q. Wang, C. Zhou, Y. Chai, *Nanoscale* **2020**, *12*, 8109.

[27] J. Park, S. Kim, M. Yang, H. Hosono, K. Park, J. Yoon, J. Bak, B. You, S.-W. Park, M. G. Hahm, M. Lee, *ACS Photonics* **2023**, *10*, 2930.

[28] C. Zhou, S. Zhang, Z. Lv, Z. Ma, C. Yu, Z. Feng, M. Chan, *npj 2D Materials and Applications* **2020**, *4*, DOI 10.1038/s41699-020-00179-9.

[29] C. Zhou, S. Raju, B. Li, M. Chan, Y. Chai, C. Y. Yang, *Advanced Functional Mater.* **2018**, *28*, DOI 10.1002/adfm.201802954.

[30] Y. Hu, J. Wang, M. Tamtaji, Y. Feng, T. W. Tang, M. Amjadian, T. Kang, M. Xu, X. Shi, D. Zhao, Y. Mi, Z. Luo, L. An, *Advanced Mater.* **2024**, DOI 10.1002/adma.202404013.

[31] J. Kang, W. Liu, D. Sarkar, D. Jena, K. Banerjee, *Physical Review X* **2014**, *4*, DOI 10.1103/physrevx.4.031005.

[32] Y. Liu, J. Guo, E. Zhu, L. Liao, S.-J. Lee, M. Ding, I. Shakir, V. Gambin, Y. Huang, X. Duan, *Nat.* **2018**, *557*, 696.

[33] W. Zhao, Z. Ghorannevis, K. K. Amara, J. R. Pang, M. Toh, X. Zhang, C. Kloc, P. H. Tan, G. Eda, *Nanoscale* **2013**, *5*, 9677.

[34] G. S. Papanai, B. K. Gupta, *Mater. Chemistry Frontiers* **2023**, *7*, 3102.

[35] Y. Li, X. Li, T. Yu, G. Yang, H. Chen, C. Zhang, Q. Feng, J. Ma, W. Liu, H. Xu, Y. Liu, X. Liu, *Nanotechnol.* **2018**, *29*, 124001.

[36] G. Xu, D. Liu, J. Li, J. Li, S. Ye, *Science China Technological Sciences* **2022**, *65*, 1263.

[37] L. Fang, H. Chen, X. Yuan, H. Huang, G. Chen, L. Li, J. Ding, J. He, S. Tao, *Nanoscale Research Letters* **2019**, *14*, DOI 10.1186/s11671-019-3110-z.

[38] X. Zhao, Z. Shi, X. Wang, H. Zou, Y. Fu, L. Zhang, *InfoMat* **2020**, *3*, 201.

[39] T. Das, S. Youn, J. E. Seo, E. Yang, J. Chang, *ACS Applied Materials & Interfaces* **2023**, *15*, 45116.

[40] Y. Wang, R. X. Yang, R. Quhe, H. Zhong, L. Cong, M. Ye, Z. Ni, Z. Song, J. Yang, J. Shi, J. Li, J. Lu, *Nanoscale* **2016**, *8*, 1179.

[41] J. Chen, Z. Zhang, J. Feng, X. Xie, A. Jian, Y. Li, H. Guo, Y. Zhu, Z. Li, J. Dong, Q. Cui, Z. Shi, C. Xu, *Advanced Mater. Interfaces* **2022**, *9*, DOI 10.1002/admi.202200075.

[42] W. Gao, S. Zhang, F. Zhang, P. Wen, L. Zhang, Y. Sun, H. Chen, Z. Zheng, M. Yang, D. Luo, N. Huo, J. Li, *Advanced Electronic Mater.* **2020**, *7*, DOI 10.1002/aelm.202000964.

[43] A. Abnavi, R. Ahmadi, H. Ghanbari, D. Akinwande, M. M. Adachi, *ACS Nano* **2024**, *18*, 34147.

[44] H. Ghanbari, A. Abnavi, R. Ahmadi, M. R. Mohammadzadeh, M. Fawzy, A. Hasani, M. M. Adachi, *Advanced Optical Mater.* **2024**, *12*, DOI 10.1002/adom.202401682.

[45] J. Yu, J. Lou, Z. Wang, S. Ji, J. Chen, M. Yu, B. Peng, Y. Hu, L. Yuan, Y. Zhang, R. Jia, *J. Of Alloys And Compounds* **2021**, *872*, 159508.

[46] D. Rauh, C. Deibel, V. Dyakonov, *Advanced Functional Mater.* **2012**, *22*, 3371.

[47] A. Abnavi, R. Ahmadi, H. Ghanbari, M. Fawzy, A. Hasani, T. D. Silva, A. M. Askar, M. R. Mohammadzadeh, F. Kabir, M. Whitwick, M. Beaudoin, S. K. O'Leary, M. M. Adachi, *Advanced Functional Mater.* **2022**, *33*, DOI 10.1002/adfm.202210619.

[48] J. Hu, Y. Shi, Z. Zhang, R. Zhi, H. Li, Y. Tang, S. Yang, B. Zou, *J. Of Mater. Chemistry C* **2019**, *7*, 14938.

[49] M. Z. Nawaz, L. Xu, X. Zhou, M. Javed, J. Wang, B. Wu, C. Wang, *ACS Applied Materials & Interfaces* **2023**, DOI 10.1021/acsami.2c22219.

[50] Z. P. Ling, R. Yang, J. W. Chai, S. J. Wang, W. S. Leong, Y. Tong, D. Lei, Q. Zhou, X. Gong, D. Z. Chi, K.-W. Ang, *Optics N.a.* **2015**, *23*, 13580.

[51] Z. Dan, B. Yang, Q. Song, J. Chen, H. Li, W. Gao, L. Huang, M. Zhang, M. Yang, Z. Zheng, N. Huo, L. Han, J. Li, *ACS Applied Materials & Interfaces* **2023**, *15*, 18101.

[52] D. Wu, J. Guo, J. Du, C. Xia, L. Zeng, Y. Tian, Z. Shi, Y. Tian, X. J. Li, Y. H. Tsang, J. Jie, *ACS Nano* **2019**, *13*, 9907.

[53] X. Tang, S. Wang, Y. Liang, D. Bai, J. Xu, Y. Wang, C. Chen, X. Liu, S. Wu, Y. Wen, D. Jiang, Z. Zhang, *Physical Chemistry Chemical Phys.* **2022**, *24*, 7323.

[54] D. Xiang, C. Han, Z. Hu, B. Lei, Y. Liu, L. Wang, W. P. Hu, W. Chen, *Small* **2015**, *11*, 4829.

[55] C. Liu, T. Zheng, K. Shu, S. Shu, Z. Lan, M. Yang, Z. Zheng, N. Huo, W. Gao, J. Li, *ACS Applied Materials & Interfaces* **2024**, *16*, 13914.

[56] K. Tang, C. Yan, X. Du, G. Rao, M. Zhang, Y. Wang, X. Wang, J. Xiong, *Advanced Optical Mater.* **2023**, *12*, DOI 10.1002/adom.202301350.

[57] B. Liu, C. Zhao, X. Chen, L. Zhang, Y. Li, H. Yan, Y. Zhang, *Superlattices and Microstructures* **2019**, *130*, 87.

[58] Y. Sun, J. Xiong, X. Wu, W. Gao, N. Huo, J. Li, *Nano Research* **2021**, *15*, 5384.

[59] W. Li, L. Liu, Q. Tao, Y. Chen, Z. Lu, L. Kong, W. Dang, W. Zhang, Z. Li, Q. Li, J. Tang, L. Ren, W. Song, X. Duan, C. Ma, Y. Xiang, L. Liao, Y. Liu, *Nano Letters* **2022**, *22*, 4429.

[60] E. Wu, D. Wu, C. Jia, Y. Wang, H. Yuan, L. Zeng, T. Xu, Z. Shi, Y. Tian, X. Li, *ACS Photonics* **2019**, *6*, 565.

[61] D. Wu, R. Tian, P. Lin, Z. Shi, X. Chen, M. Jia, Y. Tian, X. Li, L. Zeng, J. Jie, *Nano N.a.* **2022**, *104*, 107972.

[62] S. He, P. Feng, Y. Du, Y. Ma, C. Dang, A. Shan, L. Zhao, T. Wei, M. Li, L. Gao, *Advanced Optical Mater.* **2024**, *12*, DOI 10.1002/adom.202302399.

[63] J. Han, C. Fang, M. Yu, J. Cao, K. Huang, *Advanced Electronic Mater.* **2022**, *8*, DOI 10.1002/aelm.202100987.